\documentclass[12pt]{article}

\usepackage[dvipdfmx]{graphicx}

\usepackage{enumerate}
\usepackage[dvipdfmx]{color}
\usepackage{ascmac}

\usepackage{times}

\usepackage[mathlines]{lineno}

\makeatletter
\def\Hline{
\noalign{\ifnum0=`}\fi\hrule \@height 2pt \futurelet
\reserved@a\@xhline}
\makeatother

\usepackage{amsmath}
\usepackage{ascmac}
\usepackage{enumerate}
\usepackage{makeidx}

\renewcommand{\vec}[1]{\mbox{\boldmath $#1$}}

\title{
A prototype model for evaluating 
psychiatric research strategies: 
Diagnostic category-based approaches vs. the RDoC approach 
} 
\author{
Kentaro Katahira$^1$, \& Yuichi Yamashita$^2$ \\
}
\date{}
\begin{document}

\maketitle
\noindent
{$^1$ Department of Psychology, Graduate School of Environmental Studies, 
 Nagoya University, Furo-cho, Chikusa-ku, Nagoya, Aichi, Japan }\\ 

\noindent
{$^2$ Department of Functional Brain Research, National Institute of Neuroscience,
National Center of Neurology and Psychiatry, Kodaira, Tokyo, Japan}\\

\begin{abstract}
In this paper, we propose a theoretical framework 
for evaluating psychiatric research strategies. The strategies 
 to be evaluated 
 include a conventional diagnostic category-based approach 
and dimensional approach that have been encouraged by 
the National Institute for Mental Health (NIMH), outlined as 
Research Domain Criteria (RDoC). 
The proposed framework is based on the statistical modeling of the processes by which 
 pathogenetic factors are translated to behavioral measures and how the 
 research strategies can detect potential pathogenetic factors. 
The framework provides the statistical power for quantifying how 
 efficiently relevant pathogenetic factors are detected under various conditions. 
We present several theoretical and numerical results highlighting the merits 
 and demerits of the strategies. 
\end{abstract}

\noindent
Keywords: 
Research Domain Criteria; Diagnostic category; 
DSM; ICD; statistical power 

\newpage

\section{Introduction}
Psychiatry research is experiencing two major movements: 
one is the introduction 
of computational approaches \cite{montague2012computational, wang2014computational,stephan2015translational}; 
the other concerns research 
strategies in psychiatry in a more general regard, through a proposal of research 
strategy presented by the National Institute for Mental Health (NIMH), outlined as the Research Domain Criteria, or RDoC 
\cite{insel2010research, cuthbert2014rdoc, insel2014nimh}. 
While both movements appear to be promising, whether and how these movements can improve research in psychiatry is still a matter of debate. 
The main focus of the present paper is related to the second movement. 
We propose a theoretical framework for evaluating 
how research strategies including those defined by RDoC are effective 
in psychiatric research. 
However, the proposed framework also provides a basis 
on which to evaluate the contribution of the computational approaches to psychiatry research.

\subsection{Diagnostic category-based approach} 

Conventional psychiatric research aiming to find the pathogenetic factors of mental 
disorders is based on the diagnostic-category-based approach (hereafter, we simply 
refer to it as ``category-based approach'').  
Researchers classify the subjects into a clinical population 
(patient group) and non-clinical population (control group) at first. The 
classification is based on the current diagnostic systems, such as the 
Diagnostic and Statistical Manual of Mental Disorders (DSM; American 
Psychiatric Association, 2013) or the International Classification of 
Diseases (ICD; World Health Organization 1990). 
The classification is usually based on multiple criteria of symptoms or 
signs. Then, researchers attempt to determine the factors that significantly differ 
between groups. The current computational approaches to psychiatry 
are also mainly based on this category-based 
approach. For example, model parameters that are fit to the subject's 
behavior or brain activities that are correlated with model latent 
variables are compared between the control group and patient group (e.g., 
\cite{yechiam2005using, murray2008substantia, gradin2011expected, 
Kunisato2012, Ahn2014}
).  

Several methodological flaws of the conventional category-based approach 
have been pointed out (e.g., \cite{insel2010research, cuthbert2013constructing, owen2014new}). 
One notable flaw is the heterogeneity in the population
classified as the clinical population. The heterogeneity of the corresponding 
biological and social factors in a population precludes the researcher 
from detecting them. Another flaw is that similar symptoms that 
may share similar pathogenetic factors are included in different 
categories of mental disorders. For example, obsessive compulsive symptoms 
in schizophrenia are remarkably prevalent and considered as important factors in 
neurobiological studies of schizophrenia \cite{hwang2001schizophrenia, meier2014obsessive}. 
This also can obscure
determining the ultimate cause of the mental disorders. These two 
problems can be summarized as the lack of a strict one-to-one mapping from 
pathogenetic factors to the current category of the mental 
disorders; there appear to be many-to-one or one-to-many 
mappings between them. 

\subsection{Dimensional (RDoC) approach} 

To overcome the above mentioned problems, the NIMH proposed RDoC. We do not 
provide a full introduction of RDoC here (for a complete description of RDoC, 
see the RDoC website: 
http://www.nimh.nih.gov/research-priorities/rdoc/index.shtml). 
The important properties that this article discusses are as follows. RDoC 
encourage researchers to seek the relationships among the 
behavioral measurements (included as ``Behavior'' and ``Self-reports'' in the 
unit of analysis) and biological and social factors (included as ``Genes'', 
``Molecules'', ``Cells'', ``Circuits'', ``Physiology'' in the unit of analysis), 
focusing on research domains and constructs. The research domains (e.g., 
``positive valence systems'') contain constructs (e.g., ``reward learning''). 
Constructs can be subcomponents of diagnostic criteria of mental 
disorders in DSM/ICD, but the conventional categorization of the 
diagnostic systems is not used. Thus, the method of 
the analysis would be dimensional rather than categorical. 
If we assume linear relationships 
between measures in the units of analysis, a typical statistical approach is 
regression or correlation analysis. The relationship, however, is not 
necessarily linear and can be non-linear (e.g., inverted U-shaped curve; 
\cite{cuthbert2014rdoc}). Although we only focus on linear 
correlations in this article, our framework can be extended to a non-linear 
case. 

\subsection{Goal of this study}

It seems that researchers in psychiatry largely appreciate the RDoC as  
promising research strategies. However, is this indeed the case? 
Although there are methodological flaws in the current diagnostic systems 
(DSM/ICD) as discussed above, the DSM/ICD also provide advantages. One 
advantage is that the reliability of the diagnosis can be increased by using  
multiple criteria. This may lead to an increase in the 
likelihood that a researcher finds the pathogenetic factors of the mental 
disorders, compared to the RDoC approach, which decomposes the 
criteria used in the DSM into distinct dimensions. Therefore, it is 
important to clarify under what conditions the RDoC approach supersedes the 
conventional, category-based approaches. For this purpose, 
mathematical and computational models may provide a useful framework for 
addressing such questions quantitatively. The present study proposes a 
prototype for such a framework. 

\section{Proposed model} 

Here, we formally describe the proposed model. 
We assume that there are $N$ potential pathogenetic factors that can be 
causes of mental disorders. 
The $j$-th pathogenetic factor is denoted 
as $x_j$. All the pathogenetic factors are summarized as a column vector: 
$X = (x_1,...,x_N)^T$, where $\cdot^T$ denotes the transpose. 
The pathogenetic factors may include specific alleles or brain 
connectivity, which can be predictors of risk. They may also 
include the dysregulation of the neuromodulator or neurotransmitter, 
which can be a target of medical treatment, as well as the social environment 
or personal experience. 
One goal of basic research in psychiatry is to find pathogenetic factors 
that are relevant to mental disorders. 
In general, the measurement of the value of $X$ is often contaminated by 
noise that may be caused by the estimation error or measurement error. 
The measured or estimated value of $x_i$ is denoted as $\hat{x}_i$. 

Behavioral measures, including symptoms and signs that are used in 
DSM/ICD-based classification, are denoted as $Y = (y_1,...,y_M)^T$. 
In the RDoC framework, such 
behavioral measures are included in the units of analysis 
``Behavior'', or ``Self-reports''. 
Here, we consider $M$ such behavioral measures. 

\subsection{Mapping from pathogenetic factors to behavioral measures} 

The pathogenetic factors $X$ are assumed to be translated to behavioral measures 
$Y$ via some function $f$ 
with some added noise $\epsilon$. In vector form, this can be written as 

\begin{align}
Y = f ( X ) + \vec{\epsilon}, 
\end{align} 
where $\vec{\epsilon}$ is an $M$-dimensional column vector. 
$f(\cdot)$ represents a map from an $N$-dimensional column vector to an 
$M$-dimensional column vector. 
We refer to this model as a {\it generative model}. 
The noise may include the individual difference in resilience, any 
other personality trait that affects how easily the individual experiences the 
disorder, or the errors in the subjective report and behavioral measure. 

In the following analysis and simulations, we only consider a 
simple, linear and Gaussian case. 
The noise 
$\epsilon = (\epsilon_1, ..., \epsilon_M)^T$ is assumed to independently obey a 
Gaussian distribution with zero mean and common variance $\sigma_\epsilon^2$: 
\begin{align}
\epsilon_i \sim {\cal N}(0, \sigma_\epsilon^2) \ \forall_i , 
\end{align} 
where ${\cal N}(\mu, \sigma^2)$ indicates the Gaussian distribution with 
mean $\mu$ and variance $\sigma^2$. 
We also assume that the function $f$ is a linear transformation: 

\begin{align}
f ( X ) &= W X, 
\end{align} 
where $W$ is an $M \times N$ matrix. 

Furthermore, the pathogenetic factors are assumed to independently 
obey a Gaussian distribution with zero mean and unit variance: 
\begin{align} 
x_j \sim {\cal N}(0, 1) \ \forall_j.  
\end{align}

From the above assumptions, each behavioral measure, $y_i$,
marginally obeys the Gaussian distribution.
This means that behavioral measures are continuous variables.
On the other hand, many inclusion criteria in the current diagnostic systems
(i.e., DSM and ICD) take on 
discrete values (e.g., existence or absence of a symptom), with the exception of 
the duration quantity that indicates how long an episode continues for.
Thus, in this case, $y_i$ may be interpreted as a 
behavioral phenotype, based on which a psychiatrist or a patient makes decisions 
regarding each symptom, rather than the criterion itself. 

For simplicity of analysis, 
the weight parameters and noise $\epsilon$ are re-parametrized so that 
the marginal distribution of each behavioral measure, $y_j$, has unit 
variance (for details, see Appendix~\ref{appendix:normalization}). By this 
parametrization, the same fraction of individuals are classified as 
patients in category-based approach, given a set of inclusion criteria. 
For example, if there is a single criterion and the threshold is $h = 0.5$ (see below for 
the definition of $h$), approximately 31 \% of the 
individuals are classified into the clinical population on average. 
This re-parametrization does not influence the results of 
the dimensional approach. 

\subsection{Category-based approach}  
\begin{figure}[h]
 \begin{center}
 \includegraphics[width=0.8\linewidth]{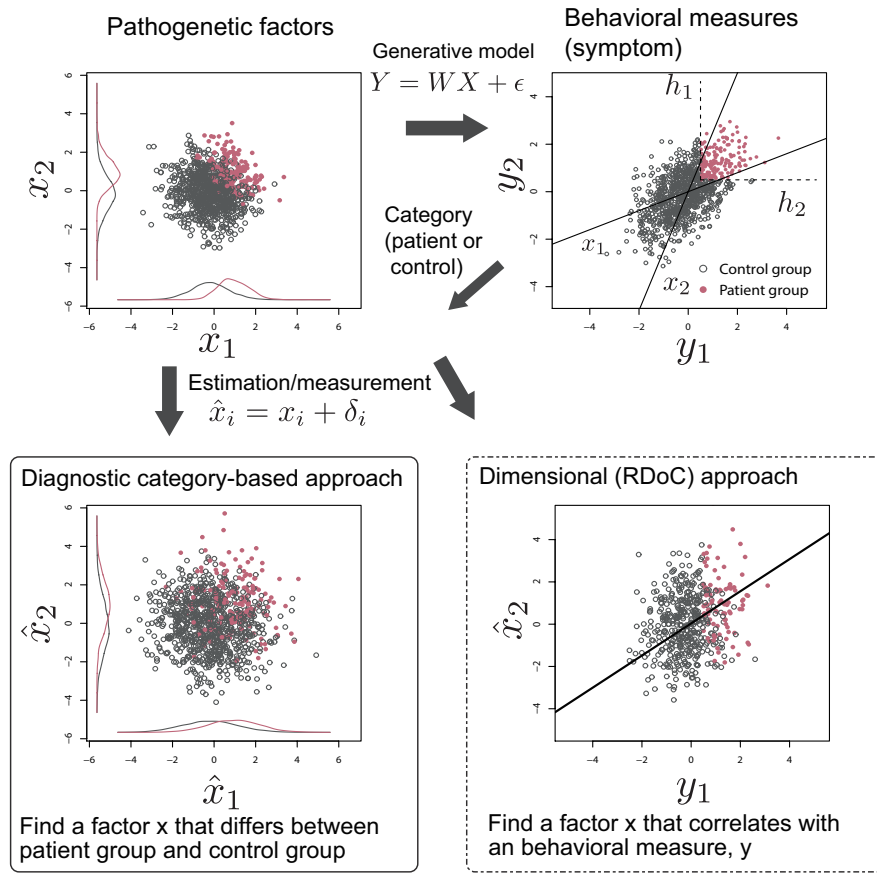}
 \end{center}
 \caption{
Illustration of the proposed model. Each dot 
 represents an individual subject. The samples were 
 generated by the model under a linear, 
 Gaussian case with $N = 2$, $M = 2$, and $c = 0.4$ (in Equation~\ref{eq:W_c}). 
The individuals are classified as patients (clinical) if both behavioral measures 
 $y_1$ and $y_2$ have larger values than $h_1$ and $h_2$, respectively (here, we used 
 the common criterion: $h_1 = h_2 = 0.5$). 
The red dots represent the patient group, and the gray dots represent the control 
 group. 
} 
 \label{fig:model_schematic}
\end{figure}

In the proposed model, the category-based approach first classifies the subjects 
into the patient group or control group depending on the values of their 
behavioral measure, $Y$. 
For example, if $y_i$ for all $i$ exceeds the threshold $h_i$ 
($y_i \ge h_i \ \forall_i$), 
the subject is classified into the patient group 
(in Figure~\ref{fig:model_schematic}, the subjects indicated with red dots 
belong to the patient group). 
Except for Case 1, where we examine the effect of the margin between the patient group and 
control group, 
the subjects who do not 
satisfy the inclusion criteria ($y_i < h_i \  \exists_i$) are classified into the 
control group (the subjects indicated with gray dots). 
In the following simulations, we set $h_i = h = 0.5 \ \forall_i$. 

The category-based approach seeks the component of $X$ that 
significantly differs between two groups. 
The estimated or measured $X$ that contains a measurement error is assumed to 
be generated by 
\begin{align*}
\hat{x}_j = x_j+ \delta_j, \ \delta_i \sim {\cal N}(0, \sigma_\delta^2). 
\end{align*}
Note that we formally and explicitly model the measurement or estimation 
error by using a Gaussian variable, $\delta_i$, rather than incorporating 
a specific estimation process. 

In the simulation, the samples of subjects ($n_1$ subjects from the control 
group and $n_2$ subjects from the patient group, resulting in $n_1 + n_2 = 
n$ subjects) are randomly selected from 
both groups, and their $\hat{x}_j$ values are subjected to an unpaired 
t-test with the equal variance assumption. If the significance of the 
difference is detected at the significance level $\alpha = .01$, 
the factor $x_i$ is deemed as a factor relevant to the mental disorder. 
When multiple candidate factors are submitted to statistical 
testing, a correction should be made for multiple comparisons (e.g., 
Bonferroni correction) to suppress family-wise error rates. However,  for 
simplicity, we do not perform the correction in this paper.  
Incorporating a correction is straightforward and does not influence the 
qualitative results reported in this paper. 

\subsection{RDoC (dimensional) approach}  

In the proposed framework, the RDoC approach is simulated by sampling 
$n$ subjects irrespective of the behavioral phenotype (symptom). 
The statistical hypothesis test is then conducted with the null 
hypothesis, where the correlation coefficient between $y_i$ and $\hat{x}_j$ 
is zero. When the correlation is significant (the null hypothesis is 
rejected), the factor $x_i$ is deemed as a factor relevant to the 
behavioral measure, $y_i$. 

\section{Results}
Below, we provide analytical and numerical results to clarify the 
basic properties of the proposed model. We especially focus on the 
statistical power, which is the probability that the pathogenetic factors are detected by 
the statistical hypothesis tests. 

\subsection{Case 1: Category-based vs. Dimensional approaches} 

\begin{figure}[h]
 \begin{center}
 \includegraphics[width=0.9\linewidth]{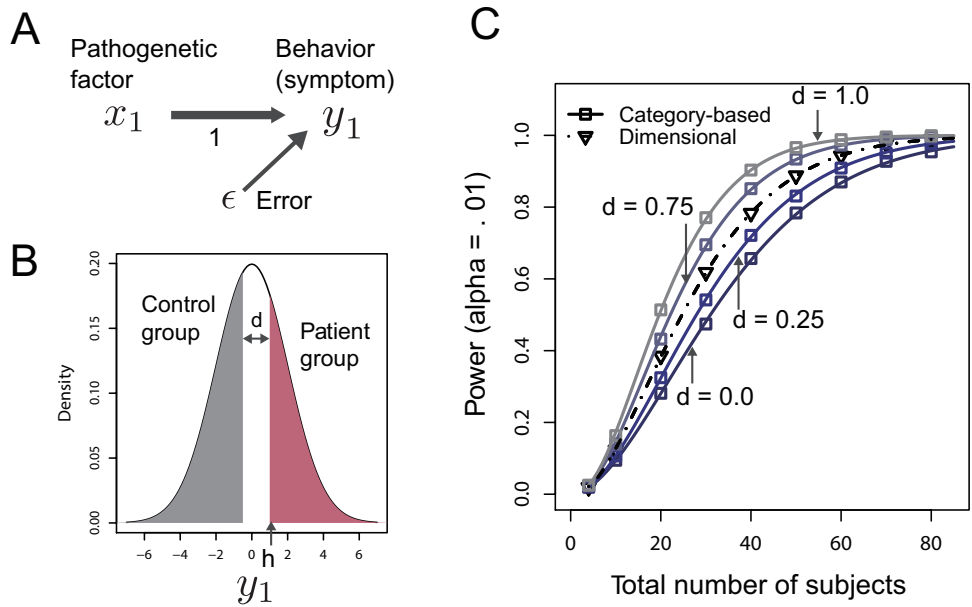}
 \end{center}
 \caption{
Comparison of the statistical power 
of the category-based approach and dimensional approaches in Case 1. 
(A) The schematic of the generative model in Case 1. 
This case includes a single pathogenetic factor ($N = 1$) and single 
 behavioral measure ($M = 1$). 
(B) Illustration of the category-based approach with a margin. 
See the main text for details. 
(C) The statistical power (with the significance level $\alpha = .01$ ) of both methods as a function of the total 
 number of subjects, with variable margin $d$ for the category-based 
 approach. The solid lines represent the analytical results 
 (see Appendix~\ref{appendix:power}). Symbols represent the results of the Monte 
 Carlo simulations (see Appendix~\ref{appendix:simulation_procedure}). 
} 
 \label{fig:categorical_vs_dimensional}
\end{figure}

First, we compare the statistical powers of the category-based approach and 
the dimensional approach for the simplest case in which there is a 
single pathogenetic factor ($M = 1$) and single behavioral measure ($N = 1$). 
The transformation matrix is a scalar, identical map: $W = 1$ (with the 
re-parametrization given in Appendix~\ref{appendix:normalization}). 
The model structure is illustrated in 
Figure~\ref{fig:categorical_vs_dimensional}A. 
We also examine the effect of the margin (denoted by $d$) 
between the patient group and control group. 
The subjects with $y$ less than $h - d$ are 
classified into the control group, while the subjects with 
$y$ larger than $h$ are classified into the patient group 
(Figure~\ref{fig:categorical_vs_dimensional}B). 
The subjects with $y$ falling into the 
margin are not included in the study. 
Actual samples in psychiatry studies may include such margins either 
intentionally or unintentionally; the researcher may exclude the subjects who 
are not classified into the clinical group  
but present behavioral phenotypes that are close to the inclusion criteria.  

Figure~\ref{fig:categorical_vs_dimensional}C shows the power that the 
pathogenetic factor $x_1$ is detected as a function of the total number 
of subjects. 
For this case, the statistical power of both methods is 
analytically obtained (see Appendix~\ref{appendix:power}; 
Figure~\ref{fig:categorical_vs_dimensional}C, lines). 
The symbols (squares for the category-based approach and triangles for the dimensional 
approach) represent the numerical results obtained from 10,000 runs of 
the Monte Carlo simulations. The results of the analysis (lines) perfectly agree 
with those obtained from the simulations, validating the analysis in 
Appendix~\ref{appendix:power}. 

The results indicate that if there is no margin ($d = 0$),  the 
dimensional approach (using correlation 
coefficients) yields a higher power compared to the category-based approach 
(using the unpaired t-test). This is because the correlation coefficients can utilize 
full information on the magnitude of $x_1$, while the 
category-based approach ignores the information of the distribution within the group. 
If there is a margin, the category-based approach can supersede the 
dimensional approach. With a larger margin, the category-based approach can distinguish 
 clusters in the distribution $x_1$ while suppressing the impact of the 
noise added to $x_1$. 
It should be noted, however, 
that with a larger margin, it becomes more difficult to find samples for 
the control group. 

\subsection{Case 2: The effect of the number of diagnosis criteria in the category-based 
  approach}
\begin{figure}[t]
 \begin{center}
 \includegraphics[width=0.8\linewidth]{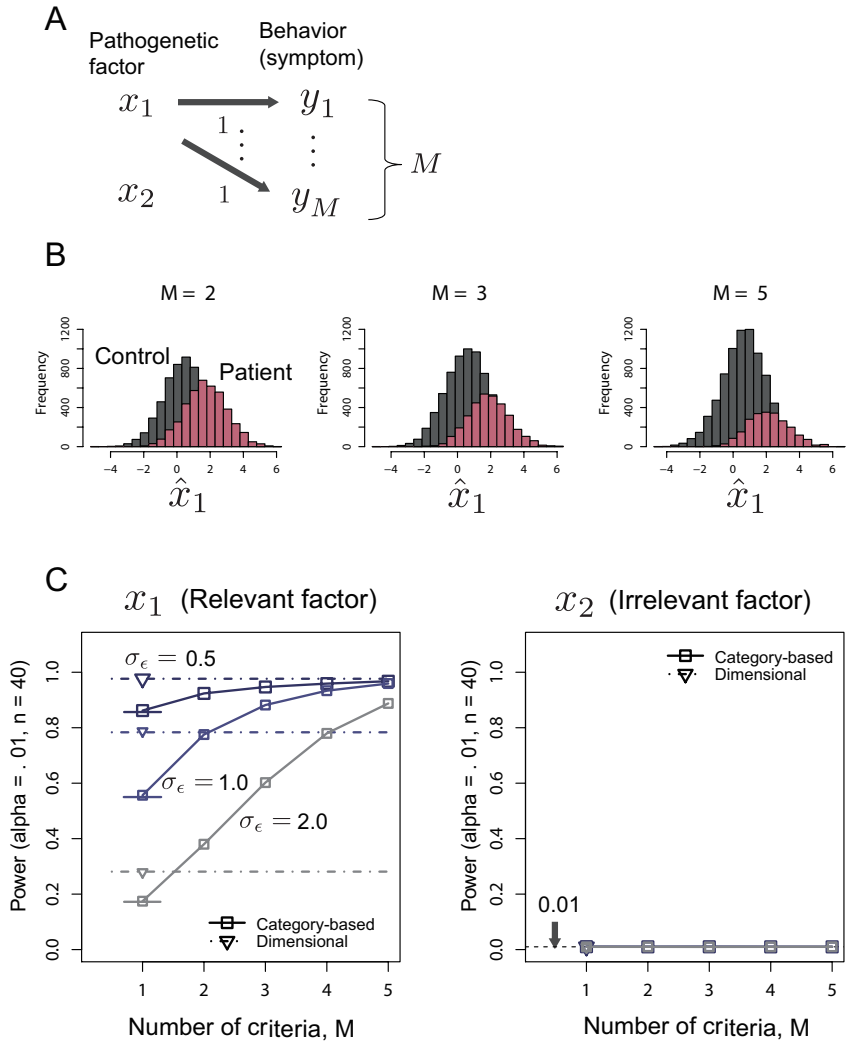} 
 \end{center}
 \caption{
The effect of the number of diagnosis criteria, $M$, in the category-based 
  approach (Case 2). 
(A) The schematic of the generative model in Case 2. 
Here, the model includes two pathogenetic factors ($N = 2$; $x_2$ is 
 irrelevant) and $M$ behavioral measures. 
(B) The distribution of the estimated pathogenetic factor $\hat{x}_1$ for 
 three $M$ cases. 
(C) The statistical power (with significance level $\alpha = .01$ ) of both 
 methods as a function of $M$, with varying standard deviation of the noise, $\sigma_\epsilon$. 
The horizontal lines at $M = 1$ represent the analytical results 
 (see Appendix~\ref{appendix:power}). The symbols and the lines 
 connecting the symbols 
 for $M$ for the category-based approach represent the results of Monte 
 Carlo simulations. 
} 
 \label{fig:effectM}
\end{figure}

As we discussed in the Introduction, the category-based approach may increase the 
reliability of the diagnosis by using multiple criteria. The following results 
illustrate this point. 
In Case 2, there are two pathogenetic factors ($N = 2$): $x_1$ is a 
factor relevant to the mental disorder and is of interest. 
$x_2$ is irreverent to the mental disorder (Figure~\ref{fig:effectM}A). 
The weight of $x_1$ for behavioral measure $y_j, (j = 1,...,M)$ is set to 
1 and that of $x_2$ is set to zero. 
When $ M = 3$, the generative model becomes 
\begin{align*}
y_1 &= x_1 + \epsilon_1, \\
y_2 &= x_1 + \epsilon_2, \\
y_3 &= x_1 + \epsilon_3. 
\end{align*}
For the vector and matrix form of the model, see 
Appendix~\ref{appendix:simulation_procedure}. 
Figure~\ref{fig:effectM}A illustrates the structure of the generative model. 

The standard deviation of the noise, $\sigma_\epsilon$, and $M$ were varied in the simulations. 
As an example the histogram of $\hat{x}_1$ shown in 
Figure~\ref{fig:effectM}B, 
the larger is the number of the criteria $M$, the 
lower is $\hat{x}_1$ of the patient group that overlaps with that of the control group. 
Consequently, as $M$ increases, the power increases 
(Figure~\ref{fig:effectM}C, left). 
The power can exceed that of the dimensional approach in which a single 
behavioral measure is used in each statistical test. 

For the irrelevant factor $x_2$, the fraction of the factor deemed 
significant was kept to the preset significance level, 0.01 
(Figure~\ref{fig:effectM}C, right). 

\subsection{Case 3: The effect of a mixture of pathogenetic factors}

We now discuss the case where a single behavioral measure $y_i$ is 
affected by more than one pathogenetic factor, $x_j$. 
It is conceivable that a larger degree of mixture leads to difficulty 
in detecting each pathogenetic factor. 
For simplicity, we consider the case with two pathogenetic factors, $N = 2$, and 
two behavioral measures, $M = 2$. 

The transformation matrix is parametrized using a parameter $c$ that 
represents the degree of the mixture (Figure~\ref{fig:effect_c}A; also see 
Equation~\ref{eq:W_c} in Appendix~\ref{appendix:simulation_procedure}). 
The generative model in element-wise form is 
\begin{align*}
y_1 = x_1 + c x_2 + \epsilon_1, \\
y_2 = c x_1 + x_2 + \epsilon_2, 
\end{align*} 
with 
$ 0 \le c \le 1$. 
When $c = 1$, $x_1$ and $x_2$ equally contribute to both behavioral 
measures, $y_1$ and $y_2$ (complete mixture). When $c = 0$, $x_1$ and $x_2$ independently 
contribute to $y_1$ and $y_2$, respectively (no mixture). 
The effect of $c$ on the transformation is illustrated in 
Figure~\ref{fig:effect_c}B. 

We consider two cases in the category-based  approach: one uses only a 
single behavioral measure $y_1$ 
as a criterion, and the other uses both behavioral measures. 
The resulting statistical powers are plotted in Figure~\ref{fig:effect_c}C. 
As the degree of the mixture, $c$, increases, the power for the methods using a 
single criterion (dimensional approach and category-based  approach using a 
single criterion) decreases. This is because the other 
pathogenetic factor functioned as noise in terms of detecting target $x_j$ 
when $c$ had a non-zero value. 
On the other hand, the power of the category-based approach using two criteria did not 
change or even increase as $c$ increased. 
The reason for this is as follows. This approach equally uses 
 $y_1$ and $y_2$. $c$ does not largely change the total information extracted 
 from $y_1$ and $y_2$. The increase in the power is due to the noise 
 reduction effect reported in Case 2. 

The additional pathogenetic factor $x_2$ is added to $y_1$ when $c$ is non-zero; 
thus, $x_2$ is detected as a relevant pathogenetic factor even when the single criterion 
$y_1$ is used (Figure~\ref{fig:effect_c}C, right panel). 

\subsection{Case 4: The effect of the number of pathogenetic factors} 

The effect of the mixture reported in Case 3 was not drastic because there 
were only two pathogenetic factors ($N = 2$). 
As the next simulation shows, when $N$ is large, the effect is 
large: it is more difficult to detect the individual pathogenetic factor 
$x_i$. 
We varied $N$ and fixed the number of behavioral criteria to 
$M = 1$. The mixture parameter $c$ was also 
included (Figure~\ref{fig:effect_N}A; 
Equation~\ref{eq:W_case4} in Appendix~\ref{appendix:simulation_procedure}
). 

The results are shown in Figure~\ref{fig:effect_N}B. 
Overall, the influence of the number of pathogenetic 
factors ($N$) and the degree of the mixture $c$ is similar for both the 
category-based and dimensional approaches. 
When the degree of the
mixture is maximum ($c = 1$), the statistical power drastically 
decreases as the number of pathogenetic factors increases. 
This decreases is modest when the degree of the mixture is small (e.g., $c = 0.3$). 
Of course, when there is no mixture ($c = 0$), the statistical power 
does not depend on the number of pathogenetic factors (data not shown). 

\begin{figure}[t]
 \begin{center}
 \includegraphics[width=0.9\linewidth]{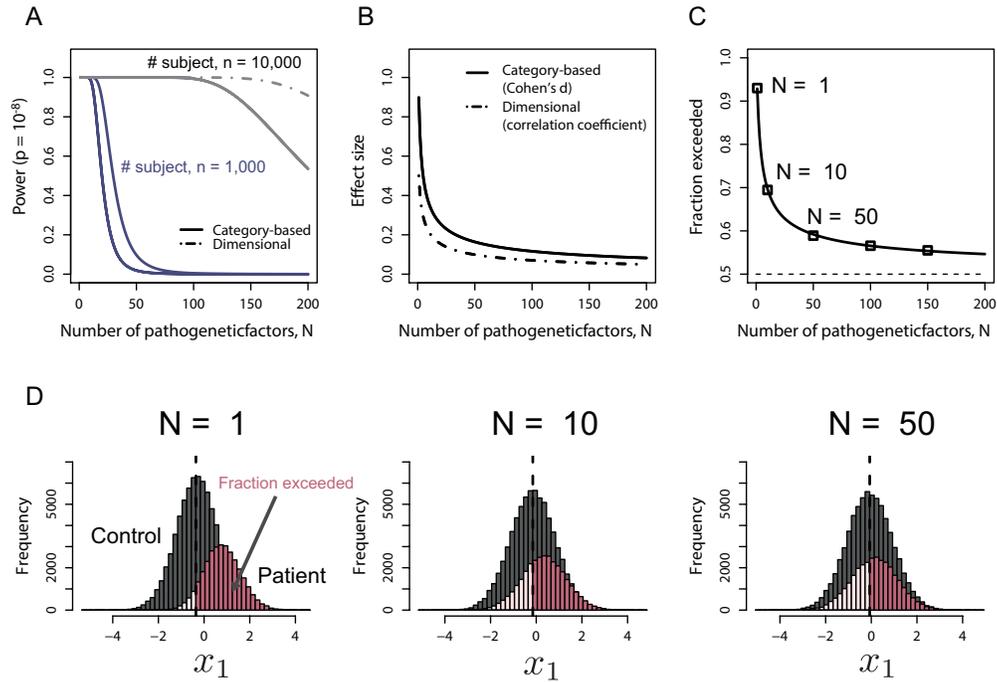}
 \end{center}
 \caption{
The effect of the number of pathogenetic factors, $N$, in the large sample case. 
(A) The statistical power (with the critical value $\alpha = 10^{-8}$ ) as a 
 function of $N$. 
(B) The effect size as a function of $N$. 
The effect size for the dimensional approach is the correlation 
coefficient. The effect size for the category-based approach is Cohen's $d$. 
(C) The fraction exceeded as a function of $N$. 
The fraction exceeded is defined as the fraction of the patients whose 
pathogenetic factor $x_1$ exceeds the mean $x_1$ of the control group, 
 as illustrated in (D). 
The lines are obtained from  the analytical results. The squares denote the
numerically obtained fraction with a total subject size of 100,000. 
} 
 \label{fig:effect_size}
\end{figure}

A large-scale psychiatry study such as genome-wide analysis (GWAS) uses larger sample sizes and, accordingly, more stringent statistical criteria. 
For example, more than 1 million alleles from about 
30,000 individuals (for both the patient group and control group) are analyzed in 
\cite{cross2013identification}. With such a large sample 
size, the factors that have very small effects on the disorder could be 
deemed as statistically significant. 
We simulated a very large sample with a stringent 
statistical criterion ($p < 10^{-8}$). 
The model structure is the same as that in Figure~\ref{fig:effect_N}A. 
The number of behavioral measures was set to $M = 1$, and 
the degree of the mixture was set to $c = 1$ (i.e., all the relevant 
pathogenetic factors equally contributed to the disorder). 
Figure~\ref{fig:effect_size} presents the results. 
When the total sample size is $n = 10,000$, even with such a stringent criterion, 
the factor $x_1$ was detected with large statistical 
power close to probability 1 (Figure~\ref{fig:effect_size}A). 
On the other hand, the effect of each pathogenetic factor drastically 
decreased as the number of relevant factors, $N$, increased (Figure~\ref{fig:effect_size}B). 
The effect size for the dimensional approach is measured by the correlation 
coefficient $\rho$ given in Equation~\ref{eq:rho_x_hat_y} in 
Appendix~\ref{appendix:power}. This $\rho$ is less than 0.2 if 
$N$ is larger than 10. 
The effect size for the category-based approach is the difference in the 
means divided by the standard deviation. This corresponds to 
the effect size called Cohen's $d$ (see Appendix~\ref{appendix:power}). 
Cohen's $d$ also easily fell below 0.2 as $N$ increased. 

To gain more insight into the effect, we computed the 
 fraction exceeded by computing the fraction of the patients for whom the 
pathogenetic factor $x_1$ exceeded the mean $x_1$ of the control group 
(Figure~\ref{fig:effect_size}D). When the fraction exceeded is 0.5 and 
the distribution is symmetric, the pathogenetic factor is irrelevant to 
the disorder. 
Figure~\ref{fig:effect_size}C plots the fraction exceeded as a function 
of $N$. 
When $N$ is greater than 50, the fraction exceeded is less than 60 \%, 
indicating that the fraction of patients who have a higher value for the pathogenetic factor $x_1$ than the healthy controls are only 10 \% above the chance level. 
For such situations, the treatment for the pathogenetic factor may have 
a limited impact. 

\section{Discussion}

In this article, we proposed a simple model for discussing the 
effectiveness of research strategies in psychiatry. 
We intended to propose this model as a basic prototype for more realistic 
applications, rather than as a model for specific psychiatric disorders. 
Thus, there are many differences between the model assumptions and realistic 
situations. Before discussing the 
discrepancies between the assumptions and the realistic situations, we discuss the implications 
derived from the analysis of the model properties. 

\subsection{Implications}

The results highlighted the effectiveness of isolating a behavioral 
measure directly associated with a pathogenetic factor. If a 
behavioral measure includes contributions from many pathogenetic 
factors, they may function as noise and reduce the chance of finding 
each relevant pathogenetic factor. Thus, the RDoC approach that decomposes the 
factors and measures 
into constructs and the unit of analysis would be promising in this 
regard. 
On the other hand, the behavioral measure can be contaminated with noise, including 
errors in the subjective report, individual differences in resilience, 
 and estimation errors in the model parameters. 
For example, the parameter estimates from the model fit to behavioral data can be used as 
behavioral measures \cite{yechiam2005using, Kunisato2012, huys2013mapping, Ahn2014}. 
However, the estimator can take on an extreme (erroneous) value. Such noise also 
prevents the researcher from detecting the factor. 
The errors may be smaller for the criteria adopted in DSM or ICD 
compared to the model estimation. 
In addition, we showed that increasing the number of independent criteria can reduce 
the impact of such noise and make the detection of the pathogenetic 
factors easier (Figure~\ref{fig:effectM}), given that the errors are mutually  
independent. 

Therefore, in some cases, the conventional diagnostic category-based 
approach could be more
efficient in detecting a pathogenetic factor than the dimensional 
(RDoC) approach: which approach is better is decided on a case-by-case basis. 
Researchers should consider these issues. 
The proposed model provides a promising way for designing an efficient research 
strategy to investigate a specific target. 

\subsection{Limitations and  possible extensions} 

We discuss the limitations of the results and possible extensions 
of the proposed framework that go beyond the limitations. 

\subsubsection{Assumptions about the model variables}

The present model assumes that the variables take continuous values and obey 
a Gaussian distribution. While this assumption makes the theoretical 
analysis easier, it is an obvious over-simplification. 
For example, consider a genetic mutation as a pathogenetic factor. 
The presence or absence of an allele is 
represented as a categorical variable. The behavioral measure or 
symptom can also be categorical (e.g., the existence or absence of a specific symptom). 
For such cases, the translation of pathogenetic 
factors to behavioral measures may be better represented as a logistic 
function. Additionally, the distribution of scores for some symptom ratings can be 
best explained using an exponential distribution with a cut-off \cite{melzer2002common}. 
The use of the link function that maps variables onto the exponential 
function with a shift parameter may be suitable for such cases. 
Although the basic properties reported in this study may hold in 
various other situations, a careful investigation would be needed depending on the situation. 

Another drastic simplification in the present model is the 
assumption of independence among errors and also among pathogenetic factors. In 
realistic situations, there may be considerable correlations among them. 
A second-order correlation can be modeled using a multi-variate Gaussian 
distribution, which is a simple extension of the current model. However,  
there may be a higher-order interactions 
among pathogenetic factors. 
Such a 
correlation structure should be included in the model depending on the specific 
situation, especially for discussing the impact of the relationships between the 
pathogenetic factors. 

\subsubsection{Mapping from the pathogenetic factor to the behavioral phenotype}
We only considered a linear transformation for the mapping from the 
pathogenetic factor $X$ to the behavioral measure $Y$. In reality, this mapping
can be highly non-linear and probabilistic. 
We certainly desire this mapping to reflect reality. 
However, in many situations, it is hard to determine the exact form of the transformation. 
Computational modeling studies may provide an explicit form of the mapping. 
For example, neural network models that can generate 
schizophrenia-like deficits provide a map of the neural 
connections and 
resulting neural activities onto the behavioral phenotypes \cite{yamashita2012spontaneous}. 
Additionally, a neural circuit model at the biophysical level 
can serve such a purpose \cite{wang2014computational}. 

The variables of computational models are often associated with 
neuromodulators \cite{angela2005uncertainty, dayan2008serotonin, 
stephan2015translational}. If there are indeed such associations, 
a model parameter or a latent variable can be used as an estimate of a 
pathogenetic factor. Computational models, including reinforcement 
learning 
models and Bayesian models, can be used to represent the translations from such 
factors to behaviors. Connecting the computational models to 
statistical models that explicitly describe behavioral tendencies would 
provide an 
efficient way of explicitly representing the transformation 
(e.g., \cite{katahira2015relation}). 
Thus, computational approaches will help connect the biological 
(neural) factor to the behavior, within the subconstructs of the RDoC. 
These approaches indicate the affinity of 
computational approaches 
for the RDoC approach. 

\subsubsection{Overlap of the pathogenetic factor in multiple disorders} 
In this article, we have considered cases with a single disorder. 
However, the co-occurrence of multiple 
disorders (i.e., comorbidity) within individuals was often observed in DSM- or ICD-based 
diagnoses. Additionally, the same factors (e.g., genetic mutation) may 
influence more than one disorder \cite{cross2013identification}. 

In a simple form, the proposed model may represent such 
situations with the following assumptions. 
Suppose there are three behavioral measures ($M = 3$) and four pathogenetic 
factors ($N = 4$). A subject is diagnosed to have disorder A if $y_1 > h$ and 
$y_3 > h$. Additionally, she or he is also diagnosed to have disorder B if $y_2 > h$ 
and $y_3 > h$. 
Behavioral measure $y_3$ represents the common symptom criteria between two 
disorders, and $y_1$ and $y_2$ are specific criteria for each disorder. 
The pathogenetic factor $x_4$ is common to both disorders, 
while $x_1$, $x_2$, $x_3$ are specific factors for each symptom. For example, this 
relation is represented  by the following generative model, 
\begin{align*}
y_1 &= w_{1} x_1 + w_{4} x_4 + \epsilon_1, \\
y_2 &= w_{2} x_2 + w_{4} x_4 + \epsilon_2, \\
y_3 &= w_{3} x_3 + \epsilon_3. 
\end{align*}
The overlap of a pathogenetic factor between diagnostic categories 
occurs via two routes. In one route, the factor indeed affects the distinct symptom in two 
disorders. In this case, $x_4$ corresponds to such a factor (with 
non-zero $w_4$). 
In the other route, due to the common symptom, $y_3$, $x_3$ corresponds to the pathogenetic factor shared by 
two disorder categories. 
The approach solely based on a diagnostic category cannot distinguish 
between these cases. This fact represents an advantage of the RDoC approach. 
In a more complicated form, the effectiveness of a 
psychiatric research strategy is more difficult to evaluate if there are 
multiple disorders that are shared with multiple pathogenetic factors. 
A systematic evaluation based on the proposed model would be 
useful for such situations.

\subsubsection{Cluster structure in the population}

We have assumed that the pathogenetic factors, the errors, are distributed 
continuously over the population. Several computational approaches attempt to find 
sub-cluster structures within the patient groups using machine learning 
methods 
\cite{brodersen2014dissecting, stephan2015translational, dayan2015taming}. 
The framework in the present paper can be extend to such a situation if 
the pathogenetic factors are assumed to be generated by a mixture of distributions. 
There may be subgroups in the mapping from the pathogenetic factor 
to a behavioral phenotype.  
For example, there may be subpopulations whose behavior can be easily 
affected by a pathogenetic factor, while 
the behavior of others is unaffected by the factor (e.g., resilience).  
Although resilience can be modeled as an error, $\epsilon_i$, there may 
be a case where it is better explained by a subcluster in the 
mapping $f$. 

\subsubsection{Research dynamics} 
The proposed model captures a single phase of a psychiatry study. 
The optimal research strategy may change depending on the progress in 
research. For example, at the beginning stage, an exploratory 
strategy would be suitable. As the candidates of the pathogenetic factor 
are narrowed down, a more detailed and careful strategy may be desirable. 
Including the dynamics of the research progress is a promising 
extension of the proposed framework. 

\subsubsection{Predictive validity}

The primary focus of the present study was the probability that the 
researcher finds a pathogenetic 
factor relevant to the disorders. The framework is extended to 
discuss the predictive validity, i.e., 
predictions of the disease process or outcome and response to the treatment. 
To discuss the predictive validity, additional assumptions 
are required, e.g., how the treatment affects the value of the 
pathogenetic factor or how the disorder progresses. 

\subsubsection{Designing novel diagnostic criteria} 
The scope of the present study is basic research strategies in psychiatry, rather than 
 clinical use. Thus, the proposed model is not intended to provide a 
diagnostic criterion, as the current RDoC is not. 
However, based on the proposed model, one can study how to optimize the 
diagnostic criteria and resulting diagnostic category. The optimization may be done so that mappings from 
pathogenetic factors to behavioral phenotypes do not have mixtures (i.e., 
so that they have a one-to-one mapping). Such an optimized diagnosis may help provide more effective treatment. The present model (or mode advanced models based on it) would 
be a useful tool for designing such new diagnostic categories. 

\section{Conclusion}

Psychiatry targets extremely complex processes, i.e., mental 
processes or mental states. 
There are many factors that influence them. 
Accordingly, there should be various research strategies in psychiatry, 
as well as in neuroscience and psychology. 
A quantitative evaluation 
of the research strategies is required. 
Discussion at the verbal description level is limited because the 
target system is very complex and may not be fully described verbally. 
Thus, computational and 
mathematical models could play important roles. 
Although there is plenty of room for modification, the present study is a 
first step towards such 
theoretical evaluations. 
Our study also provide an avenue via computational approaches for 
contributions to psychiatric research. 

\section*{Acknowledgments} 
This work was supported in part by the Grants-in-Aid for Scientific 
Research (KAKENHI) grants 26118506, 15K12140, and 25330301. 

\bibliography{C:/HOME/katahira/tex/katahira.bib} 

\newpage
\section*{Appendix}

\appendix

\section{Standardization of the behavioral measure}
\label{appendix:normalization}

Each behavioral measure $y_i$ is normalized so that the 
marginal distribution of the population distribution (rather than the sample 
distribution) has zero mean and unit variance. 
$y_i$ can be written as $y_i = \sum_{k}^N w_{ik} x_k + \sigma_\epsilon 
z_i$, where $z_i$ is a random variable with zero mean and unit variance 
and $x_k$ and $z_i$ are independent. 
In general, the variance of the sum of independent random variables can 
be written as follows: for $y = a x_1 + b x_2$ where $x_1$ and $x_2$ are 
independent random variables, $\text{Var}(y) = a^2 \text{Var}(x_1) + b^2 \text{Var}(x_2) $. 
By this relation, the variance of the marginal distribution of the behavioral 
measure $y_i$ is $\sum_{k}^N w_{ik}^2 + \sigma_\epsilon^2$. Thus, 
with reparametrization: 
\begin{align} 
a_{i} &= \sqrt{\sum_{k}^N w_{ik}^2 + \sigma_\epsilon^2}, \\
y_i &\leftarrow y_i / a_i \ \forall_i, 
\end{align}
the marginal distribution of $y_i$ obeys a Gaussian distribution with 
zero mean and unit variance. 
Here, the component in the $i$-th row and the $j$-th column of $W$ is denoted as
$w_{ij}$. 
Equivalently, this normalization can be achieved with the following 
parameterization: 
\begin{align} 
w_{ij} &\leftarrow w_{ij}/a_{i}, \\
\epsilon_{i} &\leftarrow \epsilon_{i}/a_{i}. 
\end{align}
In the main text and the Appendix, we used the parametrized forms of
$w_{ij}$ and $\epsilon_{i}$. 

\section{Power analysis}
\label{appendix:power}

Here, we analytically derive the statistical power, 
the probability of the correct rejection of the null 
hypothesis, for the case $M = 1$ for both the category-based analysis and correlation analysis. 
The formulation we consider here is summarized as 
\begin{align*}
y & =  \frac{\sum_{k}^N w_{k} x_k + \epsilon}
{\sqrt{\sum_{k}^N w_{k}^2 + \sigma_\epsilon^2}}, \\
\hat{x}_j & =  x_j + \delta,
\end{align*}
with the random variables obeying Gaussian distributions: 
\begin{align*}
x_j & \sim  {\cal N}(0, 1) \ \forall_j, \\
\epsilon & \sim {\cal N}(0, \sigma_\epsilon^2), \\
\delta & \sim {\cal N}(0, \sigma_\delta^2). 
\end{align*}
Here, we omitted the subscript for the index of $y$. 
The variances of $y$, $\hat{x}_j$ are respectively 
\begin{align}
\text{var}(y^2) &= 1, 
\ 
\text{var}(
\hat{x}_i^2
) = 1 + \sigma_\delta^2 . 
\end{align}
The covariance between $y$ and $\hat{x}_j$ is 
\begin{align}
\text{cov} (y, \hat{x}_j) = \frac{w_i}
{\sqrt{\sum_{i=1}^N w_i^2 + \sigma_\epsilon^2}}. 
\end{align}
The correlation coefficient between two variables ($x$, $y$) is given by 
\begin{align}
\rho_{x, y} = 
\frac{\text{cov}(x, y)}
{\sqrt{\text{var} (x)}
\sqrt{\text{var} (y) }}. 
\end{align}
Thus, the correlation coefficients between $x_j$ and $y$ and between 
$\hat{x}_j$ and $y$ are given by 

\begin{align}
\rho_{x_j, y} = \frac{w_j}
{\sqrt{\sum_{k=1}^N w_k^2 + \sigma_\epsilon^2}}, \ 
\rho_{\hat{x}_j, y} = 
\frac{w_j}
{\sqrt{\sum_{k=1}^N w_k^2 + \sigma_\epsilon^2} \sqrt{1 + \sigma_\delta^2}}, 
\label{eq:rho_x_hat_y}
\end{align}
respectively. 

\subsection{Category-based approach}
To derive the statistical power of the category-based approach, 
we first calculate the mean and variance for each group. 
The mean and variance of $x_j$ given the condition, $y > h $, i.e.,  the case 
where the subject is classified as a patient, are (cf., \cite{minotani2012seiki}):

\begin{align}
\text{E} [x_j| y > h] & = \rho_{x_j, y} \lambda(\alpha_1), \\ 
\text{var}[x_j| y > h] & = 
 1 - \rho_{x_j, y}^2 \lambda(\alpha_1) \left[
\lambda(\alpha_1) - \alpha_1 
\right], 
\end{align}
where 
\begin{align}
\lambda(\alpha_1) & = \frac{\phi(\alpha_1)}{ 1 - \Phi(\alpha_1)}, \\
\alpha_1 & = \frac{h}{\text{var}(y)} = h 
\end{align}
and 
\begin{align}
\phi(x) &= \frac{1}{\sqrt{2 \pi}} \exp \left( - \frac{1}{2} x^2 \right), \\
\Phi(x) &= \int_{-\infty}^x \phi(u) d u. 
\end{align}
Accordingly, the mean and variance of $\hat{x}_j$ are given by 
\begin{align}
\text{E} [\hat{x}_j| y > h] & = 
\rho_{\hat{x}_j, y} \lambda(\alpha_1), \\ 
\text{Var}[\hat{x}_j| y > h] & = 
1 - \rho_{\hat{x}_j, y}^2 
\lambda(\alpha_1) \left[\lambda(\alpha_1) - \alpha_1 
\right] + \sigma_\delta^2. 
\end{align}
Similarly, given $y < h - d $ (the subject is classified into the control group), 
the mean and variance of $\hat{x}$ are given by 

\begin{align}
\text{E} [\hat{x}_j| y < h - d] & = - \rho_{\hat{x}_j, y} \lambda(\alpha_2), \\ 
\text{var}[\hat{x}_j| y < h - d] & = 
1 - \rho_{\hat{x}_j, y}^2 \lambda(\alpha_2) \left[
\lambda(\alpha_2) - \alpha_2 
\right] + \sigma_\delta^2, 
\end{align}
where 
\begin{align}
\alpha_2 & = - \frac{h - d}{\text{var}(y)} = d - h. 
\end{align}

Using these expressions, the effect size of
the difference between two groups can be obtained. 
Here, we consider the effect size defined as the difference in means ($\mu_1 - \mu_2$) divided 
by the standard deviation ($\sigma$) for each mean. We assume that the t-test can be 
performed with the assumption that the variance is common to both groups. 
Although this is not actually the case, when $h$ is not so far from zero, this 
is good approximation. Additionally, the t-test is known to be robust against the 
difference in the variance between two groups (it is known that if the ratio 
 of the standard deviation is lower than approximately 1.5, the violation of the 
 assumption does not influence the result). 

Specifically, the effect size considered here is given by
\begin{align}
d_{\text{eff}} = \frac{ \mu_1 - \mu_2}{\sigma}. 
\end{align}
This is the special case of Cohen's $d$ if the number of samples for 
both groups is the same ($n_1 = n_2 = n/2$). 
The population means ($\mu_1$, $\mu_2$) and the common standard 
deviation, $\sigma$, are given by 
\begin{align}
\mu_1 & =  \rho_{\hat{x}_j, y} \lambda(\alpha_1), \\
\mu_2 & =  - \rho_{\hat{x}_j, y} \lambda(\alpha_2), \\ 
\sigma & = 
\sqrt{
 1 - 
\frac{\rho_{\hat{x}_j, y}^2 }{2} 
\left\{ 
\lambda(\alpha_1) \left[
\lambda(\alpha_1) - \alpha_1 
\right]
+ 
\lambda(\alpha_2) \left[
\lambda(\alpha_2) - \alpha_2 
\right]
\right\} 
+ \sigma_\delta^2
}.  
\end{align}
Here, the means of the variance of both groups were used to approximate the common variance. 

The test statistic used for the t-test is 
\begin{align}
t = d_{\text{eff}} \times \sqrt{\frac{n_1 n_2}{n_1 + n_2}}. 
\end{align}
The test statistic $t$ obeys the Student's t-distribution with a degree of freedom $n_1 + n_2 - 2$ under the null assumption. 
When the alternative hypothesis ($H_1$: $\mu_1 \ne \mu_2$) is correct, 
$t$ obeys a 
noncentral t-distribution with a degree of freedom $n_1 + n_2 - 2$ and noncentrality parameter given by 
\begin{align}
\lambda = 
d_{\text{eff}} \times \sqrt{\frac{n_1 n_2}{n_1 + n_2}}. 
\end{align}
Using this fact, the statistical power can be obtained by using the 
following R code: 
\begin{verbatim}
tcritical <- qt(1-pcritical/2, df = n1 + n2 - 2) 
power <- pt(-tcritical, df = n1 + n2 - 2, 
            ncp = eff_size * sqrt(n1*n2/(n1+n2)))  + 
            1 - pt(tcritical, df = n1 + n2 - 2, 
            ncp = eff_size * sqrt(n1*n2/(n1+n2))) 
\end{verbatim}
Here, the meaning of the variables is as follows; 
\verb|pcritical|: significance level, $\alpha$; 
\verb|eff_size|: effect size, $d_{\text{eff}}$; 
\verb|n1|: number of samples in the patient group; 
\verb|n2|: number of samples in the control group. 

\subsection{Dimensional approach} 
Next, we consider the test for correlation between $\hat{x}_j$ and $y$ 
where the null hypothesis is $\rho_{\hat{x}_j, y} = 0$. 
The test statistic is 
\begin{align}
t = \frac{r_{\hat{x}_j, y}}{\sqrt{1 - r_{\hat{x}_j, y}^2}} \times \sqrt{ 
 n - 2}, 
\end{align}
where $r_{\hat{x}_j, y}$ is the sample correlation coefficient between 
the $n$-samples of 
$\hat{x}_j$ and $y$. 
Under the null assumption, $t$ obeys the Student's t-distribution with  
degree of freedom: $n - 2$. 

The statistical power for this case can be obtained by using the ``pwr'' package 
in R that uses the approximation proposed in  
\cite{cohen1988statistical}. 
Specifically, the following R code is used: 
\begin{verbatim}
pwr.r.test(n = n, r = rho, sig.level = 0.01) 
\end{verbatim}
Here, the meaning of the variables is as follows; 
\verb|n|: number of total samples; 
\verb|rho|: (true) correlation coefficient, $\rho_{\hat{x}_i, y}$, given in 
Equation~\ref{eq:rho_x_hat_y}. 

\section{Simulation procedure} 
\label{appendix:simulation_procedure}
All the simulations and numerical calculations presented in this paper 
were performed in R version 3.2.0 \cite{R2015}. 
The details of the simulation settings for each case are described below. 
Common settings for the model parameters are  
$\sigma_\delta = 1$, $\sigma_\epsilon = 1$, unless
otherwise specified. 

The Gaussian random variables in the models are sampled from the ``rnorm'' 
function in R. We sampled the data for 10,000 individuals for each run. 
To obtain the statistical power numerically, the simulation was run  
100,000 times for each condition, and we count the fraction where a factor was 
deemed as significant. 

The data generation was performed based on matrix multiplication. 
Examples of the matrix and vector forms are provided below. 
For Case 2, where $ M = 3$ , the vectors and matrix become  
\begin{align}
Y = 
\left(
\begin{array}{c}
y_1 \\
y_2 \\
y_3 
\end{array}
\right), 
\ 
W = \left(
 \begin{array}{cc}
 1 & 0 \\
1 & 0 \\
1 & 0 
\end{array}
\right),
X = 
\left(
\begin{array}{c}
x_1 \\
x_2 
\end{array}
\right). 
\end{align}
For Case 3, the transformation matrix $W$ becomes 
\begin{align}
W = \left(
 \begin{array}{cc}
 1 & c \\
 c & 1
\end{array}
\right), 
\label{eq:W_c}
\end{align}
with 
$ 0 \le c \le 1$. 
For Case 4, when $N = 4$, the transformation matrix $W$ becomes a row vector: 
\begin{align}
W = &
\left(
 \begin{array}{cccc}
 1 & c & c & c
\end{array}
\right), 
\label{eq:W_case4}
\end{align}
with $ 0 \le c \le 1$. 

\newpage

\section*{Figures}

\begin{figure}[t]
 \begin{center}
  \includegraphics[width=0.8\linewidth]{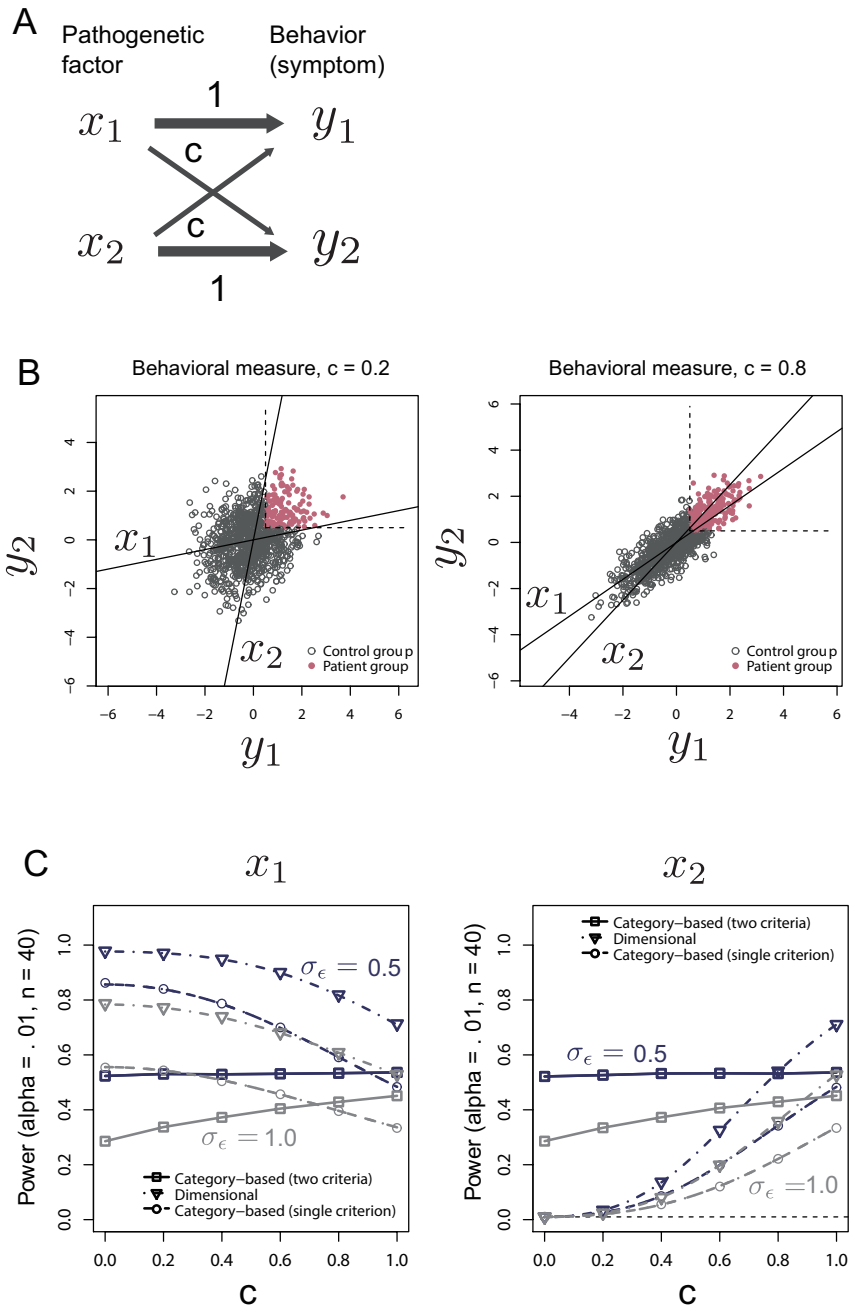}
 \end{center}
 \caption{
The effect of a mixture of pathogenetic factors (Case 3). 
(A) The schematic of the generative model in Case 3. 
Here, the model includes two pathogenetic factors ($N = 2$) and two 
 behavioral measures ($M = 2$). The parameter $c$ indicates the degree of the mixture. 
(B) The scatter plot of $Y$ for two $c$ cases. 
(C) The statistical power (with critical value $\alpha = .01$ ) of both 
 methods as a function of $c$, with varying standard deviation of the noise, $\sigma_\epsilon$. 
The dash-dot lines for the dimensional approach and the dashed lines for the 
 category-based approach with a single criterion denote the analytical results 
 (see Appendix~\ref{appendix:power}). 
Symbols and solid lines for the category-based approach using two criteria represent the 
 results of the Monte Carlo simulations. 
} 
 \label{fig:effect_c}
\end{figure}

\begin{figure}[t]
 \begin{center}
 \includegraphics[width=0.9\linewidth]{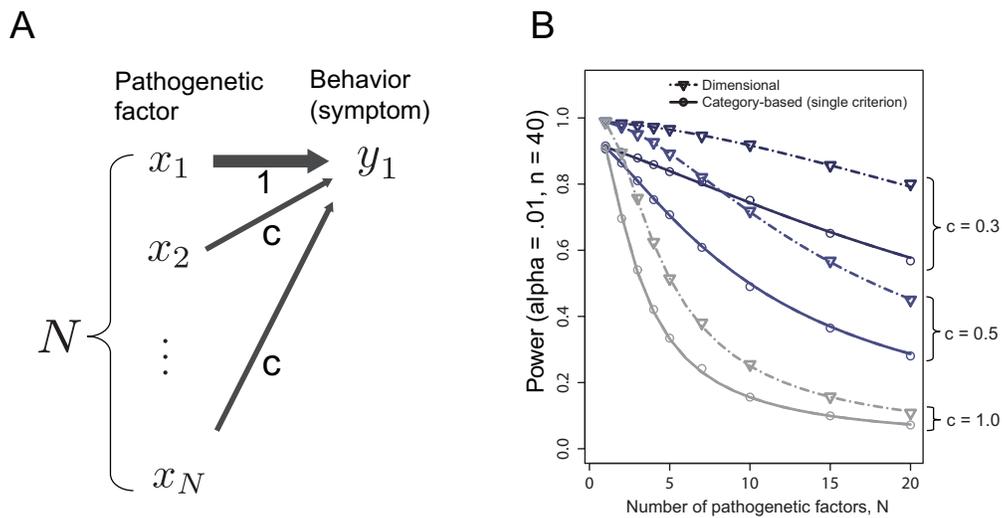} 
 \end{center}
 \caption{
The effect of the number of pathogenetic factors, $N$. 
(A) The schematic of the generative model in Case 4. 
The model includes $N$ pathogenetic factors and one 
 behavioral measure ($M = 1$). 
(B) The statistical power (with critical value $\alpha = .01$ ) of both 
 methods as a function of $N$. 
The dash-dot lines and solid lines denote the analytical results. 
Symbols represent the results of the Monte Carlo simulations. 
} 
 \label{fig:effect_N}
\end{figure}

\end{document}